\documentclass[aps,prd,showpacs,twocolumn,floats]{revtex4}
\usepackage{graphicx}
\usepackage{subfigure}
\usepackage{epsfig}
\usepackage{dcolumn}
\usepackage{bm}
\usepackage[ansinew]{inputenc}
\usepackage{amsmath}
\usepackage{amsthm}
\usepackage[T1]{fontenc}
\usepackage{amssymb}
\usepackage{amsfonts}
\usepackage[english]{babel}
\usepackage{enumitem}

\newcommand{\ds}{\displaystyle}

\begin{document}

\title{A Model of Dust-like Spherically Symmetric Gravitational Collapse without Event Horizon Formation}

\author{Miquel Pi\~{n}ol}
 \altaffiliation {Unitat de Medicina Intensiva, Hospital Universitari i Polit\`{e}cnic La Fe. Val\`{e}ncia. Spain.}


\date{\today}

\begin{abstract}
Some dynamical aspects of gravitational collapse are explored in this paper. A time-dependent spherically symmetric metric is proposed and the 
corresponding Einstein field equations are derived. An ultrarelativistic dust-like stress-momentum tensor is considered to obtain analytical 
solutions of these equations, with the perfect fluid consisting of two purely radial fluxes - the inwards flux of collapsing matter and the 
outwards flux of thermally emitted radiation. Thermal emission is calculated by means of a simplistic but illustrative model of uninteracting 
colllapsing shells. Our results show an asymptotic approach to a maximal space-time deformation without the formation of event horizons. The 
size of the body is slightly larger than the Schwarzschild radius during most of its lifetime, so that there is no contradiction with either 
observations or previous theorems on black holes. The relation of the latter with our results is scrutinized in detail.
\end{abstract}


\maketitle

\section{\label{introduction}Introduction}

The aim of this paper is to discuss several open problems of conceptual interest concerning black holes and, in particular, to elaborate a 
simple model of dust-like spherically symmetric gravitational collapse with account of both the inwards flux of the collapsing matter and 
the outwards flux of emitted thermal radiation. We illustrate how the latter, usually ignored in models of gravitational collapse (except 
in some few references, see \cite{Sto1,Sto2}), may avoid the formation of event horizons. The metric considered in this work is time-dependent, 
unlike the Schwarzschild one. 
Spherical polar coordinates will be used and there will be no need for analytical extensions (such as the one given by the Kruskal-Szekeres chart) 
because the occurrence of an event horizon at the Schwarzschild radius will be avoided. 

In Sec. \ref{history} the main historical events concerning the development of the well-known concept of black hole are reviewed and its 
precise significance is shortly but precisely detailed. In Sec. \ref{open} some open problems of the common black hole model are pointed out  
and their relationship with the corresponding historical findings is emphasized. Section \ref{sec:metric} deals with the development of the metric 
of the present model: First of all, in subsec. \ref{einsteinfieldequations} a time-dependent spherically symmetric metric in spherical polar 
coordinates is presented and the corresponding Einstein field equations are specified. Secondly,  a dust-like energy momentum tensor for a purely 
radial motion with account of an ultrarelativistic collapsing matter and thermally emitted radiation is obtained in subsec. \ref{stressmomentum}. 
Temporal evolution of the metric components is studied in subsec. \ref{temporal}, with the absence of emitted thermal radiation being detailed 
as a particular case. Fourthly, in subsec. \ref{stability} it is showed that there should exist a limit where the inwards flux of collapsing matter 
and the outwards flux of thermal radiation become \emph{compensated}. It is also showed the asymptotic character of the approximation to this limit. 
Some additional considerations about the total mass and the edge of the collapsing body will be made in subsec. \ref{totalmassedge}. Finally, 
our results are discussed in Sec. \ref{Discussion}, paying a special attention to the plausibility of the different hypothesis and the implications 
of their alternatives.

\section{\label{history}Important Historical Results concerning Black Holes}

Several historical results in General Relativity led to the concept of black hole. The following list includes the most important ones: 
\begin{itemize}
\item[$1.$] K. Schwarzschild found in 1916 an exact solution of the Einstein field equations describing the field created by a point particle \cite{Sch}. 
According to Birkhoff's theorem, this solution is also valid for any spherically symmetric body at a distance larger than its radius \cite{Bir}. 
\item[$2.$] J. R. Oppenheimer and G. Volkoff discovered in 1939 the existence of upper limit for the mass of neutron stars, above which gravitational 
collapse could not be avoided \cite{Opp}. 

\item[$3.$] In 1967 J. Wheeler coined the term 'black hole' to name a 'gravitationally completely collapsed star' \cite{Whe}. 

\item[$4.$] S. Hawking and R. Penrose proved in 1970 that, under certain circumstances, singularities could not be avoided. This is known as the 
Hawking-Penrose theorem of singularity \cite{Haw1}. 
\end{itemize}
All these results concerning black holes arise basically from Einstein's General Relativity. On the other hand, there exist two important features in the 
description of black holes which require from both Thermodynamics and Quantum Field Theory (QFT):
\begin{itemize}
\item[$5.$] J. Bekenstein defined the entropy of black holes in 1972  and, based on thermodynamic grounds, deduced the need for black-hole radiation \cite{Bek}. 

\item[$6.$] In 1974 S. Hawking justified Bekenstein's speculations about the existence of black-hole radiation from the point of view of QFT. Hawking 
model implies the creation of particles of negative mass near the event horizon of black holes. The conservation of information is not clearly 
ensured by this model \cite{Haw2}. 
\end{itemize}

\section{\label{open}Some Open Problems in Gravitational Collapse}

In this section we discuss if the previous historical results genuinely imply the actual existence of black holes as physical objects. It is widely 
believed that these findings prove the existence of black holes. The argument supporting black hole formation is the following:

\begin{enumerate}
\item There exist stars which are massive enough to exceed the Oppenheimer-Volkoff limit at the end of their 'vital cycle'. Those stars must 
finally enter collapse.
\item According to the Hawking-Penrose theorem of singularity, all the mass inside an event horizon must reach a single central point, that is, 
form a singularity.
\item The solution of the Einstein field equations for the metric of a 'point mass' is the Schwarzschild metric, which describes a black hole.
\end{enumerate}

Entering collapse, however, does not immediately lead to the formation of an event horizon and, while the event horizon is not formed, the 
Hawking-Penrose theorem of singularity is not properly applicable (notice that one of its conditions of application is equivalent either to the 
existence of an event horizon, or to an expanding Universe taken as a whole). Hence, \emph{a priori} entering collapse must not necessarily lead to a 
complete collapse. As a matter of fact, Hawking has recently defended the incompatibility of event horizons with Quantum Mechanics \cite{Arx}.

Certainly, the period of time involved in the process of collapse may be proved to be infinite from the point of view of any external 
observer (that is, from our perspective on Earth). On the other hand, a 'free falling observer' would measure a finite period of time for the 
collapse, at least if nothing destroys it before reaching its goal \cite{Pen}. A well-known feature of General Relativity is that space and time are 
relative but events are absolute. Consequently, it is necessary to reconcile the observations from both reference frames.

It is usually assumed that the free falling observer actually reaches the singularity in a finite time, and the infinite-lasting collapse 
measured by the external observer is justified in the following way: the free falling body has already reached the central singularity, but as 
the light emitted from the body inside the black hole never escapes from it, we cannot see it falling; furthermore, the light emitted near the 
event horizon of the black hole comes to us with a great delay, making us believe that it is still falling.

In fact, there are compelling reasons that make us doubt about the previous explanation:
The Schwarzschild metric is symmetric under temporal inversion, which suggests that trajectories in the corresponding space-time 
should be reversible, in contrast to the most common interpretation of black holes and their event horizon. Furthermore, General Relativity is 
not only intended to explain what an observer 'sees' in a given reference frame, but what truly 'occurs' in there.

Solution of this apparent paradox requires a careful analysis of what an external observer would exactly see when looking at a body free falling
towards a black hole. On the one hand, it would see the free-falling body approaching asymptotically to the event horizon of the black 
hole, without ever crossing it. On the other hand, according to Hawking's law of black hole radiation, the observer should also see the whole black hole 
evaporating in a very large, but finite period of time. The evaporation of the whole mass of the black hole must logically include that of the 
free-falling body as well. Were it not to be like this, that is, if the crossing of the event horizon had to be accomplished before the emission of thermal 
radiation, it would never emit thermal radiation and the laws of Thermodynamics would be infringed. As the temporal order of causally-related events is 
always the same for all reference frames, we must conclude that the free falling observer should also observe its own complete evaporation before having 
reached the event horizon. If it had reached the singularity in a finite period of time, its complete evaporation must have occured in a finite and 
\emph{lesser} period of time.

Not only should these considerations be valid for the free-falling body approaching a black hole, but also for the process of collapse itself. 
Consequently, collapsing bodies should never becomes black holes. On the contrary, they should asymptotically tend to form an event horizon until 
the time at which they become completely emitted in the form of radiation. An equivalent thesis has already been defended by Vachaspati, Stojkovic and Krauss \cite{Sto1,Sto2}, and by Pi\~{n}ol and L\'{o}pez-Aylagas \cite{Pin}. Previously, Mitra had already established the mathematical inconsistency of event horizon formation in General Relativity and proposed the concept of 'Eternal Collapsing Object' \cite{Mitra1, Mitra2, Mitra3, Mitra4}. Similarly, also Robertson and Leiter developed consistent models of gravitational collapse without event horizon formation, which also take into account the magnetic effects of the supposedly identified 'black holes' \cite{Robertson1, Robertson2, Robertson3}. Additionally, there exist some calculations in string theory which point towards the same direction \cite{Kar}.

Thus, the metric of a collapsing body shall never be strictly the Schwarzschild's one (as it never completely collapses) but a time-dependent 
metric. In the next section, we solve the Einstein field equations of a time-dependent spherically symmetric metric. Several simplifications are 
considered to make calculations plausible, but the essential Physics of the problem is respected.

\section{\label{sec:metric} Deduction of a Metric for Gravitational Collapse}

\subsection{\label{einsteinfieldequations} Einstein Field Equations}

As we have already pointed out, our goal in this paper is to study the temporal evolution of a spherically symmetric gravitational collapse. 
Rotations and local inhomogeneities are beyond the scope of the present work. Therefore, the starting point shall be a time-dependent spherically 
symmetric metric, which in spherical polar coordinates is given by the expression

\begin{equation}
d\tau^2 = e^{\nu} dt^2 - e^{\lambda} dr^2 - r^2 d\Omega^2 ,
\label{metric}
\end{equation}
where $\nu=\nu(r,t)$ and $\lambda=\lambda(r,t)$. Notice that geometrized units have been used ($G=1,c=1$). The corresponding Einstein field 
equations for such metric are the following \cite{Lan}:

\begin{equation}
8\pi T^{0}_{0} = - e^{-\lambda} \left({1\over {r^2}} - {{\lambda'} 
\over {r}} \right) + {{1} \over {r^2}} , 
\label{einstein00}
\end{equation}

\begin{equation}
8\pi T^{1}_{1} = - e^{-\lambda} \left( {{\nu'} \over {r}}+ {{1} \over {r^2}} \right) + {{1} \over {r^2}} ,
\label{einstein11}
\end{equation}

\begin{align}
8\pi T^{2}_{2}= &- {1 \over 2} e^{-\lambda} \left( \nu'' + {{\nu'^2} \over {2}} + {{\nu'-\lambda'} \over {r}} + {{\nu'\lambda'} \over {2}} 
\right)\\
 &+ {1 \over 2} e^{-\nu} \left( \ddot{\lambda} + {{\dot{\lambda}}^2 \over 2} - {{\dot{\lambda} \dot{\nu}} \over 2} \right),\nonumber
\label{einstein22}
\end{align}

\begin{equation}
8\pi T^{3}_{3} = 8\pi T^{2}_{2} ,
\label{einstein33}
\end{equation}

\begin{equation}
8\pi T^{1}_{0} = - e^{-\lambda} {\dot{\lambda} \over {r}} .
\label{einstein10}
\end{equation}

Subtraction of \eqref{einstein11} from \eqref{einstein00} yields the identity

\begin{equation}
8\pi \left( T^{0}_{0} - T^{1}_{1} \right) =  {e^{-\lambda} \over r}  \left( \nu'+ \lambda' \right)
\label{addingTT}
\end{equation}

It will be useful to define a function $\phi(r,t)$ in the following way:

\begin{equation}
-2 \phi \equiv \nu + \lambda ,
\label{phi1}
\end{equation}
so that

\begin{equation}
\nu = - \lambda - 2 \phi 
\label{phi2} 
\end{equation}
and

\begin{equation}
8\pi \left( T^{0}_{0} - T^{1}_{1} \right) =  {e^{-\lambda} \over r} \left( -2 \phi' \right).
\label{phiprima}
\end{equation}

A mathematical structure for the stress-momentum tensor must be specified in order to solve the previous equations, which will be discussed in next 
subsection.

\subsection{\label{stressmomentum} A Dust-like Stress-Momentum Tensor of Ultrarelativistic particles}.

The stress-momentum tensor of a perfect fluid may be written in terms of the energy density $\rho$, the pressure $p$ and the four-velocity 
$u^\alpha$ as:

\begin{equation}
T^{\alpha}_{\beta} = g_{\beta \delta} \left( \rho + p \right) u^{\alpha} u^{\delta} - \eta^{\alpha}_{\beta} p 
\label{smtensor_pf}
\end{equation}

If the pressure appears to be very small compared to the energy density, in the limit $p \rightarrow 0$ one obtains the stress-momentum tensor 
of dust:

\begin{equation}
T^{\alpha}_{\beta} = g_{\beta \delta} \rho u^{\alpha} u^{\delta}
\label{smtensordust}
\end{equation}

In our model we deal with a dust-like stress-momentum tensor. For the sake of simplicity, we shall consider the perfect fluid splitting into 
two perfectly radial fluxes: a flux of ingoing collapsing matter and a second flux of outgoing thermal radiation. Both the ingoing collapsing 
matter and the outgoing thermal radiation are going to be dealt as ultrarelativistic particles. It has been already established that the matter 
in a process of gravitational collapse reaches celerities near the speed of light \cite{Yan}. It is also a well-known fact that, despite photons 
being 'massless', a photon gas may be assimilated to a gas of ultrarelativistic particles with an effective mass density \cite{Haz}. 



It could be expected that the relation between pressure and mass-energy density should be given by the identity $\ds p = {\rho \over 3}$ due to the 
particles being ultrarelativitic. A closer insight into this points out that the above identity would only be properly applicable to an 
\emph{isotropic} gas and not to the higly \emph{directed} movement considered in the present work. The consideration of two purely 
'radial' fluxes shall simplify calculations and it is in this sense that a 'dust-like' stress-momentum tensor may be used. A similar 
approach has been already adopted by Borkar and Dhongle \cite{Bor}.


With account of the metric \eqref{metric} the coefficients of the dust energy-momentum tensor \eqref{smtensordust} become

\begin{eqnarray}
T^{0}_{0} &=& e^{-2\phi} e^{-\lambda} \rho \left(u^{0} \right)^2 ,
\label{sm00}\\
T^{1}_{1} &=& - e^{\lambda} \rho \left(u^{1} \right)^2 ,
\label{sm11}\\
T^{1}_{0} &=& e^{-2\phi} e^{-\lambda} \rho u^{0} u^{1} .
\label{sm10}
\end{eqnarray}

For a purely radial movement (characterized by $d\Omega = 0$) Eq. \eqref{metric} leads to the relation 
\begin{equation}
d\tau^2 = e^{-2\phi}e^{-\lambda} dt^2 - e^{\lambda} dr^2 
\label{rmotion1}
\end{equation}
which, with account of the identities $\ds {dt \over d\tau} \equiv u^0$ and $\ds {dr \over d\tau} \equiv u^1$, becomes 

\begin{equation}
1 = e^{-2\phi}e^{-\lambda} \left(u^0\right)^2 - e^{\lambda} \left(u^1\right)^2 .
\label{rmotion2}
\end{equation}

Isolating $\left|u^1\right|=\sqrt{\left(u^1\right)^2 }$, we obtain 
\begin{equation}
\left|u^1\right| = e^{-\phi}e^{-\lambda} u^0 \left[1 - { e^{2\phi}e^{\lambda} \over \left(u^0\right)^2} \right]^{1 \over 2}.
\end{equation}
In the ultrarelativistic limit $u^0 \rightarrow \infty$ ($u^0\gg e^{2\phi}e^{\lambda}$) the component $u^{1}$ of the four-velocity becomes

\begin{equation}
\left|u^1\right| = e^{-\phi}e^{-\lambda} u^0 .
\end{equation}
Notice that this same relation could have been obtained by imposing the identity $d\tau \sim 0 $ in Eq. \eqref{rmotion1}.

Concerning the sign of $u^1$, it is clear that $u^1 < 0$ for ingoing matter and $u^1 < 0$ for outgoing thermal radiation. That is,

\begin{eqnarray}
u^1_{in} &=& - e^{-\phi}e^{-\lambda} u^0 ,
\label{u1in}\\
u^1_{out}&=& e^{-\phi}e^{-\lambda} u^0 .
\label{u1out}
\end{eqnarray}

\subsubsection{Stress-Momentum Tensor of the Ingoing Matter}

If we denote the energy density of the infalling matter by $\rho_{in}$, according to Eqs. \eqref{sm00}, \eqref{sm11}, \eqref{sm10} and 
\eqref{u1in} we have 

\begin{eqnarray}
T^{0}_{0, in} &=& e^{-2\phi} e^{-\lambda} \rho_{in} \left(u^{0} \right)^2 ,\label{infalling00}\\
T^{1}_{1, in} &=& - e^{-2\phi} e^{-\lambda} \rho_{in} \left(u^{0} \right)^2 \\
&=& - T^{0}_{0, in} ,\nonumber\\
T^{1}_{0, in} &=& - e^{-3\phi} e^{-2\lambda} \rho_{in} \left(u^{0} \right)^2 \label{infalling10}\\
&=& - e^{-\phi}e^{-\lambda} T^{0}_{0, in}.\nonumber
\end{eqnarray}

\subsubsection{Stress-Momentum Tensor of the Outgoing Thermal Radiation}

Again, denoting the energy density of the outgoing thermal radiation by $\rho_{out}$, according to Eqs. \eqref{sm00}, \eqref{sm11}, \eqref{sm10} and 
\eqref{u1out} we obtain

\begin{eqnarray}
T^{0}_{0, out} &=& e^{-2\phi} e^{-\lambda} \rho_{out} \left(u^{0} \right)^2 ,\\
T^{1}_{1, out} &=& - e^{-2\phi} e^{-\lambda} \rho_{out} \left(u^{0} \right)^2\\
&=& - T^{0}_{0, out} ,\nonumber\\
T^{1}_{0, out} &=& e^{-3\phi} e^{-2\lambda} \rho_{out} \left(u^{0} \right)^2\label{outgoing10}\\
&=& e^{-\phi}e^{-\lambda} T^{0}_{0, out}  .\nonumber
\end{eqnarray}

\subsubsection{Total Stress-Momentum Tensor of the Collapsing Body}

Addition of the stress-momentum tensors of both the infalling matter and the outgoing thermal radiation leads to the total stress-momentum tensor of the 
collapsing body, which is given by the expressions

\begin{eqnarray}
T^{0}_{0} &=& e^{-2\phi} e^{-\lambda} \left( \rho_{in} + \rho_{out} \right) \left(u^{0} \right)^2 ,
\label{total00}\\
T^{1}_{1} &=& - e^{-2\phi} e^{-\lambda} \left( \rho_{in} + \rho_{out} \right) \left(u^{0} \right)^2\label{total11}\\
&=& - T^{0}_{0},\nonumber\\
T^{1}_{0} &=& - e^{-3\phi} e^{-2\lambda} \left( \rho_{in} - \rho_{out} \right) \left(u^{0} \right)^2\label{total10}\\
&=& - e^{-\phi}e^{-\lambda} \left( { {\rho_{in} - \rho_{out}} \over {\rho_{in} + \rho_{out}} } \right)  T^{0}_{0}  .\nonumber
\end{eqnarray}

Once the mathematical structure of the stress-momentum tensor of the collapsing body is established, we are able to study the temporal evolution of 
the collapse by solving the Einstein field equations \eqref{einstein00}-\eqref{einstein10}.

\subsection{\label{temporal} Studying the Temporal Evolution of Collapse}

Substitution of $T^{1}_{0}$ by Eq. \eqref{total00} in Eq. \eqref{einstein10} leads to the following equation:
\begin{equation}
- e^{-\phi}e^{-\lambda} \left( { {\rho_{in} - \rho_{out}} \over {\rho_{in} + \rho_{out}} } \right)  8\pi T^{0}_{0} = - e^{-\lambda} {\dot{\lambda} \over {r}} .
\end{equation}
From this an expression for the temporal evolution of $\lambda$ may be isolated:

\begin{equation}
\dot{\lambda} = e^{-\phi} \left( { {\rho_{in} - \rho_{out}} \over {\rho_{in} + \rho_{out}} } \right) \left( 8\pi r T^{0}_{0} \right) .
\label{telambda}
\end{equation}

Initially it is expected that $\rho_{in}\gg \rho_{out}$, as the amount of energy emitted in the form of thermal radiation should reasonably correspond to 
a very small proportion of the total energy of the collapsing body. In that case,  $\left( { {\rho_{in} - \rho_{out}} \over {\rho_{in} + 
\rho_{out}} } \right) \sim 1$ and $ \dot{\lambda} \sim e^{-\phi} \left( 8\pi r T^{0}_{0} \right)$, so that $\lambda$ shall be a strictly 
increasing function with time and it is expected to acquire considerably large values. In any case, for $\lambda\gg1$ we have the asymptotic expression 
\begin{equation}
8\pi T^{0}_{0} = {{1} \over {r^2}} + O (e^{-\lambda}).
\label{AsympT00}
\end{equation}
and therefore, 
\begin{equation}
\dot{\lambda} = e^{-\phi} \left( { {\rho_{in} - \rho_{out}} \over {\rho_{in} + \rho_{out}} } \right) {{1} \over {r}} + O (e^{-\lambda}).
\label{lambdadot}
\end{equation}

On the other hand, we need to estimate as well the value of $\phi$. From Eqs. \eqref{phiprima} and \eqref{total11} we obtain

\begin{equation}
\phi' = - {1 \over 2} e^{\lambda} 8\pi r \left( T^{0}_{0} - T^{1}_{1} \right) = - e^{\lambda} \left(8\pi r T^{0}_{0} \right) ,
\end{equation}
which combined with Eq. \eqref{AsympT00} yields
\begin{equation}
\phi' = - { e^{\lambda} \over r} + \left(\frac{1}{r}-\lambda'\right)\sim - { e^{\lambda} \over r}
\label{phiprima2}
\end{equation}

According to Birkhoff's theorem, outside the radius $R$ of the collapsing body the space-time geometry will be exactly Schwarzschild-like, 
so that $\phi = 0$ for $r > R$. Inside the collapsing body $T^0_0 > 0$ and consequently $\phi' < 0$. This yields $\phi > 0$ for $r < R$ and 
$\phi(R,t)=0$ because of the analytic character of this function.

Equations \eqref{lambdadot} and \eqref{phiprima2} are not trivial to resolve analytically. For any time $t$, however, Eq. \eqref{lambdadot} and the 
fact that $\phi > 0$ for any $r < R$ lead to the following inequality:

\begin{equation}
\lambda(t,r) < \lambda(0,r) + {t \over r}
\label{lambdainequality}
\end{equation}
   
\subsection{\label{stability} Asymptotic Approach to a Pseudo-Stability Phase}
   
According to the results obtained in the previous section, for any given time $t$ the function $\lambda(r,t)$ is analytic on the domain 
$r > 0$. Nonetheless, as Eq. \eqref{lambdainequality} is an inequality, no specific values for this function have been provided. 

It has been discussed that the ingoing flux of infalling matter is initially expected to be much larger than the outgoing flux of thermal 
radiation. Despite this, as $\lambda$ becomes larger, according to Eq. \eqref{phiprima2} $\left| \phi' \right|$ must also increase. On the 
other hand, as $\phi\geq 0$ the ingoing flux must decrease according to Eq. \eqref{infalling10}. 

As the values of $T^1_{0,in}$ may become as small as wanted, if $\lambda$ and $\phi$ were not upper bounded it would not be unreasonable 
to think that the ingoing flux of infalling matter may eventually become \emph{compensated} by the outgoing flux of thermal 
radiation. It could be discussed as well that, according to Eq. \eqref{outgoing10}, the flux of outgoing thermal radiation may also become 
arbitrarily small, but we proceed first to analyse the details concerning the compesation of fluxes and the consequences of this hypothesis.

The condition for the compensation of both fluxes is naturally given by the equation

\begin{equation}
T^1_{0, in, s} + T^1_{0, out, s} = 0
\label{compensation1}
\end{equation}

It must not be misunderstood as a transgression of Oppenheimer-Volkoff's theorem. The star \emph{is not} in equilibrium. It is actually 
collapsing, as nothing prevents the infalling matter of keeping in collapse. There would simply be an additional flux (arguable in the basis 
of thermodynamic grounds, and justifiable by the conversion of a portion of the collapsing matter into thermal radiation due to the interaction 
of their respective fields) that would compensate the energy interchange across a given surface of $r-$radius.

In that hypothetical state of 'stability', from Eqs. \eqref{infalling10} and \eqref{outgoing10} a relation between the energy densities 
$\rho_{in}$ and $\rho_{out}$ can be derived
\begin{equation}
\rho_{in, s} = \rho_{out, s} = {1 \over 2} \rho_s ,
\label{rhostability}
\end{equation}
where the subindex $s$ stands for 'stability' (notice that the aforementioned relations are specific of that hypothetical phase).

Several considerations concerning the emission of thermal radiation due to collapsing bodies must be made in order to proceed further with the theoretical 
development. 

\subsubsection{A Model of Hawking-like Radiation}

According to Hawking \cite{Haw2}, the temperature of a black hole is proportional to the inverse of its Schwarzschild radius ($R_S$) and the 
thermal radiation emission rate is proportional to the inverse of the square of $R_S$:

\begin{equation}
\dot M_{H} = - {k \over R_S^2} .
\label{Hawkingr}
\end{equation}
We have denoted the thermal emission by $\dot M_{H}$ as it implies a loss in the total mass of the black hole.

In what follows, both the approach and the nomenclature adopted in the study of the mass and its mathematical relation with the components of the 
stress-momentum tensor and with the functions $\nu(r,t)$ and $\lambda(r,t)$ of the metric \eqref{metric} are the ones given in Ref. \cite{Lan}. 
The total mass of a spherically symmetric body of radius $R$ is given by the following expression:
\begin{equation}
M = \int_{0}^{R} {4\pi {r}^2 T^0_0 (r,t) dr} .
\end{equation}
Analogously, the mass contained inside a surface of radius $r$ (concentric to the spherically symmetric body of interest) is given by
\begin{equation}
m(r,t) = \int_{0}^{r} {4\pi \tilde{r}^2 T^0_0 (\tilde r,t) d\tilde{r}} .
\label{massr}
\end{equation}

Comparing Eqs. \eqref{einstein00} and \eqref{massr}, the following relation can be set between $m(r,t)$ and $\lambda(r,t)$:
\begin{equation}
e^{-\lambda(r,t)} = 1 - {2m(r,t) \over r},
\label{lambdamass}
\end{equation}
and therefore we have
\begin{equation}
- e^{-\lambda} \dot \lambda = - {2 \dot m \over r},
\end{equation}
or equivalently,
\begin{equation}
\dot \lambda =  {2 \dot m \over r} e^{\lambda} .
\end{equation}

Despite the fact that there is solely 'one' function $\lambda(r,t)$, it is useful to split $\dot\lambda$ into the sum of $\dot \lambda_{in}$ (due to 
the ingoing flux $\dot m_{in}$ of collapsing matter) and $\dot \lambda_{out}$ (due to the outgoing flux $\dot m_{out}$ of thermal radiation). In so 
doing we obtain

\begin{equation}
\dot \lambda = \dot \lambda_{in} + \dot \lambda_{out} 
\label{lambdainout}
\end{equation}
with 
\begin{eqnarray}
\dot \lambda_{in} &= & {2 \dot m_{in} \over r} e^{\lambda} 
\label{dotlambdain}\\
\dot \lambda_{out} &=&  {2 \dot m_{out} \over r} e^{\lambda} 
\label{dotlambdaout}
\end{eqnarray}

As pointed out before, the thermal emission of black holes $\dot m_{H}$ is given by Eq. \eqref{Hawkingr}. On the other hand, Vachaspati \emph{et al.} 
showed that the thermal emission of a collapsing shell approaching the Schwarzschild's radius of a Black Hole would follow a law of the 
same style \cite{Sto1}: according to their calculations, the temperature of the collapsing shell turns out to be proportional to the Hawking's one 
($T_V\sim 2.4 T_H$, where $T_V$ stands for Vachaspati's temperature and $T_H$ for Hawking's temperature).

With account of Eq. \eqref{lambdamass} the metric \eqref{metric} becomes 

\begin{align}
d\tau^2 = \Bigg( 1 - &{2m(r,t) \over r} \Bigg) e^{-2\phi(r,t)} dt^2\label{metricmass}\\
&- \left( 1 - {2m(r,t) \over r} \right)^{-1} dr^2 - r^2 d\Omega^2 ,\nonumber
\end{align}
where the resemblance with Schwarzschild's metric results evident. Certainly, there exist two main differences between Eq. \eqref{metricmass} and 
the Schwarzschild's metric: $1)$ the mass is not a constant, but a function of the radius. $2)$ there is an additional factor $e^{-2\phi(r,t)}$ 
in the coefficient $g_{00}$.

However, if $1)$ we deal with motions whose variation in the $r$-coordinate is small enough and $2)$ we assume a temporal proximity to 
the hypothetical stationary situation that we were postulating (that is, $\dot m(r,t) \sim 0$ and $\dot \phi(r,t) \sim 0$), then the metric 
\eqref{metricmass} may be \emph{locally} transformed into the Scwarzschild's one. In fact, in the \emph{vicinity} of a given radius $R_a$, 
where $m(r,t) \sim M_a$ and $ \phi(r,t) \sim \Phi_a$, we have

\begin{equation}
d\tau^2 \sim \left( 1 - {2M_a \over r} \right) d\tilde{t}^2 - \left( 1 - {2M_a \over r} \right)^{-1} dr^2 - r^2 d\Omega^2 ,
\label{metricmassapprox}
\end{equation}
with 
\begin{equation}
d \tilde t \equiv e^{-\Phi_a} dt .
\end{equation}

At this point it is time to introduce our Hawking-like radiation model. We will conceptually split the collapsing body into a sequence of 
concentric spherical shells, each of which asymptotically approaches its corresponding radius $r=2M_a$ in the coordinate system given by the 
metric \eqref{metricmassapprox}. We assume that these collapsing shells do not interact with each other. Along the lines of Ref. \cite{Sto2} it 
can be deduced that the radiation law obtained for a spherical shell asymptotically approaching in time $t$ the event horizon of a black hole 
is also valid for any of the concentric shells asymptotically approaching in time $\tilde t$ its corresponding $r=2M_a$ radius in our model.
Consequently,
\begin{equation}
{d m_{out} \over {d \tilde t}}  = - {k \over r^2}
\label{tildetradiationmodel}
\end{equation}
and so
\begin{equation}
\dot m_{out} \equiv {d m_{out} \over {d t}} = {d \tilde t \over {d t}} {d m_{out} \over {d \tilde t}} =- e^{- \phi}{k \over r^2}.
\label{tradiationmodel}
\end{equation}
From this, we straightforwardly obtain the identity
\begin{equation}
\dot \lambda_{out} =  {2 e^{\lambda} \over r} \left({- e^{- \phi} k \over r^2} \right) = - e^{- \phi}{2k e^{\lambda} \over r^3} .
\label{lambdadot1}
\end{equation}

On the other hand, according to Eqs. \eqref{einstein10} and \eqref{infalling10} an equivalent expression for $\dot{\lambda}_{in}$ is given by 
\begin{equation}
 \dot{\lambda}_{in}=e^{-\phi}(8\pi r T_{0,in}^{0}).
\label{lambda2}
\end{equation}
From Eqs. \eqref{sm00},\eqref{infalling00} and \eqref{AsympT00} we conclude that, asymptotically,
\begin{equation}
8\pi T_{0,in}^{0}=\frac{\rho_{in}}{\rho_{in}+\rho_{out}}\frac{1}{r^{2}}+O(e^{-\lambda})
\end{equation}
and therefore, with account of Eq. \eqref{rhostability}, we obtain
\begin{equation}
\dot{\lambda_{in}} =  e^{-\phi} \left( { {\rho_{in}} \over {\rho_{in} + \rho_{out}} } \right) {{1} \over {r}} \simeq
 e^{-\phi} {1 \over 2} {{1} \over {r}}
\label{lambdadot2}
\end{equation}

The stability phase is naturally defined by the condition
\begin{equation}
\dot{\lambda_{s}} = 0 
\label{stabilitycondition}
\end{equation}
and therefore, from Eqs. \eqref{lambdainout}, \eqref{lambdadot1}, \eqref{lambdadot2} and \eqref{stabilitycondition} we obtain the relation
\begin{equation}
- e^{-\phi_s} {2k e^{\lambda_s} \over r^3} + e^{-\phi_s} {1 \over 2r} = 0.
\label{stability02} 
\end{equation}
Equivalently, 
\begin{equation}
e^{\lambda_s} = {1 \over 4k} r^2 ,
\label{stability03}
\end{equation}
from which a functional dependence of $\lambda$ on $r$ is obtained for the stability phase
\begin{equation}
\lambda_s = - \ln \left( 4k \right) + \ln \left( r^2 \right).
\label{stability03ln}
\end{equation}

Taking into account Eq. \eqref{phiprima2}, from the previous equation we easily obtain an expression for $\phi_s$:

\begin{equation}
\phi'_s = - {{e^{\lambda_s}}\over {r}} = - {{r}\over {4k}} ,
\label{phiprimas}
\end{equation}
Integration over $r$ with account of the contour condition $\phi(R,t)=0\;\forall t$ discussed in the previous section yields the identity
\begin{equation}
\phi_s \left(r\right) = \int_{R}^{r} {\phi'_s \left(\tilde r\right) d\tilde r} =  {1 \over {8k}} \left( R^2- r^2 \right) ,
\end{equation}
and thus
\begin{equation}
e^{- \phi_s \left(r\right)} = e^{{-1 \over {8k}} \left( R^2- r^2 \right)} .
\end{equation}

It must be noticed that the existence of the postulated stability phase is self-consistent and that it may be clearly derived from equations 
\eqref{dotlambdain} and \eqref{dotlambdaout}: both $\left| \dot \lambda_{in} \right|$ and $\left| \dot \lambda_{out} \right|$ decrease as 
$\phi(r,t)$ increases by a factor $e^{-\phi(r,t)}$, but only $\left| \dot \lambda_{out} \right|$ increases as $\lambda(r,t)$ increases 
(by a factor $e^{\lambda(r,t)}$). Consequently, even when initially $\left| \dot \lambda_{out} \right| \ll\left| \dot \lambda_{in} 
\right|$ at large enough times both quantities should become of the same magnitude.

Nonetheless, a significant issue concerning the behaviour of $\lambda(r,t)$ for small values of $r$ must be remarked. We are going to deal it with detail in the following subsection.

\subsubsection{\label{smallradii}Corrections to the Equation of $\lambda_s$ for Small Radii}

Due to the greater emission of thermal radiation in the inner shells -see Eq. \eqref{Hawkingr}-, the expression for $\lambda(r,t)$ obtained in Eq. \eqref{stability03ln} turns out to be strictly increasing with $r$. As 
the value of the coefficient $k$ appearing in Hawking's law is very small (because of thermal emission being actually a very slow process), 
$\lambda(r,t)\gg1$ for most of the values of $r$, which is in agreement with our initial considerations. However, for 
$r \sim 2\sqrt{k}$ the approximation in Eq. \eqref{AsympT00} fails: From Eq. \eqref{lambdamass} it becomes evident that $\lambda(r,t)\ge 0\;\forall r, t$ 
due to the positivity of $m(r,t)\; \forall r, t$. Therefore, Eq. \eqref{stability03ln} cannot be valid for $r < 2\sqrt{k}$. 

Two points need to be raised in order to understand this result:
\begin{itemize}
\item[$1.$] In the domain of spacetime where $\lambda(r,t)$ ceases to be $\gg 1$, the full expression for the component $T_{0}^{0}$ of the stress-momentum 
tensor in Eq. \eqref{einstein00} should be used instead of Eq. \eqref{AsympT00}.
\item[$2.$] The amount of energy thermally emitted by a collapsing shell cannot exceed the total energy of the shell. The divergence of Hawking's radiation 
law \eqref{Hawkingr} for very small values of $r$ requires a recalculation of the corresponding energy emission for the special case of the most internal shells. 
Nevertheless, it may be argued that the fast rate of thermal emission in the center of the collapsing body should tend to keep low energy densities 
in there, contrarily to what we could have initially expected. As a matter of fact, if $m(r,t)\sim 0$ for $r < 2\sqrt{k}$, then $\lambda(r,t)\sim 0$ 
and, according to Eq.  \eqref{einstein00}, also $T^0_0 \sim 0$ holds.
\end{itemize}

On the other hand, both the function $\lambda(r,t)$ and the component $T_{0}^{0}$ of the stress-momentum could be calculated with a higher degree of precision by means of 
an iterative scheme, that is, by 1) replacing the expression for $\lambda(r,t)$ into Eq. \eqref{einstein00}, 2) recalculating $\dot \lambda_{in}$ with the 
new expression for $T^0_0$, 3) imposing the stability condition \eqref{stabilitycondition}, 4) obtaining the corresponding new expression for $\lambda(r,t)$ 
and 5) repeating the whole iteration.

For instance, from Eqs. \eqref{einstein00} and \eqref{stability03ln} we obtain:

\begin{eqnarray}
8\pi T^{0}_{0} &=& {1 \over r^2} \left(1 - e^{-\lambda} \left( 1 - r \lambda' \right) \right) \sim {1 \over r^2} \left( 1 + {4k \over r^2} \right) ,
\label{cycle1}\\
\dot \lambda_{in} &=& e^{- \phi} \left( {\rho_{in} \over {\rho_{in} + \rho_{out}} } \right) \left( 8\pi r T^0_0 \right) 
\nonumber\\
&\sim& 
{e^{- \phi} \over 2r} \left( 1 + {4k \over r^2} \right),
\label{cycle2}\\
e^{\lambda_{s}} &\sim& 1 + {r^2 \over 4k},
\label{cycle3}\\
\lambda_{s} &\sim& \ln \left( 1 + {r^2 \over 4k} \right).
\label{cycle4}
\end{eqnarray}
In this first iteration, an expression for $\lambda(r,t)$ with the expected behaviour $\lambda> 0 \ \forall r$  has been already obtained. 

The complete resolution of this problem for the range of small radii is beyond the scope of this paper. Nevertheless, as pointed out before, 
it does not represent a major setback to our conclusions. 

The asymptotic characteristics of the process leading to the stability phase will be studied in the next subsection. In that study, the special case of small 
radii is not going to be considered in this paper.

\subsubsection{\label{smallvariations}Small Variations of $\lambda(r,t)$ Before the Stability Phase}

According to Eqs. \eqref{lambdainout}, \eqref{lambdadot1} and \eqref{lambdadot2} we have

\begin{equation}
\dot{\lambda}(r,t) = e^{-\phi} \left( {1 \over 2r} - {2k e^{\lambda} \over r^3} \right)  
\label{temporallambda}
\end{equation}

In the stability phase, defined by Eq. \eqref{stabilitycondition}, the functional dependence of $\lambda$ is given by Eq. \eqref{stability03}. 
Now we proceed to study small variations of $\lambda(r,t)$ before it acquires the stability value, that is,
\begin{equation}
\lambda(r,t) = \lambda_s(r) - \lambda_{\Delta}(r,t) 
\label{closetostability}
\end{equation}

Notice that, by definition, $\dot \lambda_s(r) = 0$. This fact implies
\begin{equation}
\dot \lambda(r,t) = - \dot \lambda_{\Delta}(r,t) 
\label{closetostability2}
\end{equation}
Furthermore, because of the inequality $\lambda_{\Delta}\ll\lambda$, we will consider $\phi \simeq \phi_s$. Therefore, from Eqs. \eqref{stability03}, 
\eqref{temporallambda}, \eqref{closetostability} and \eqref{closetostability2} we obtain the expression
\begin{equation}
\dot\lambda_{\Delta} = -{e^{-\phi_s}\over 2r} \left( {1- e^{-\lambda_{\Delta}}} \right)
\end{equation}
In the limit $\lambda_{\Delta} \ll 1 $ we can approximate ${1- e^{-\lambda_{\Delta}}} \sim \lambda_{\Delta}$, so that 
\begin{equation}
\dot\lambda_{\Delta} = - {e^{-\phi_s}\over 2r} {\lambda_{\Delta}} + O \left({\lambda_{\Delta}^2}\right) ,
\label{lambdadelta}
\end{equation}
whose integration over $t$ leads to the following solution
\begin{equation}
\lambda_{\Delta} = A(r) \exp \left(- {{e^{-\phi_s}\over 2r}t}\right) = A(r)\exp\left(- e^{{-1 \over {8k}} \left( R^2- r^2 \right)} 
{t\over 2r}\right) ,
\label{lambdadeltaasymptotic}
\end{equation}
where $A(r)$ is an arbitrary positive defined function depending on the initial conditions of the problem.

Therefore, according to the hypothesis of the model, $\lambda(r,t)$ asymptotically approaches its stability value:

\begin{align}
\lambda(r,t) = - \ln ( 4k& ) + \ln \left( r^2 \right) \\
&- A(r) \exp \left(- e^{{-1 \over {8k}} \left( R^2- r^2 \right)} {t\over 2r}\right) \nonumber
\label{lambdaasymptotic}
\end{align}

\subsection{\label{totalmassedge} Some Considerations about the Mass and the Edge of the Collapsing Body}

From Eq. \eqref{u1in} the infalling velocity $\dot r_{in}$ of any collapsing shell in the present model is given by
\begin{equation}
\dot r_{in} \equiv {dr \over dt} = {dr \over d\tau} {d\tau \over dt} = {u^1_{in} \over u^0} = - e^{-\phi} e^{-\lambda}
\label{collapsevelocity}
\end{equation}

According to Eqs. \eqref{lambdamass} and \eqref{collapsevelocity} and with account of the contour condition $\phi(R,t)=0\;\forall t$, the 
motion of the edge $R$ of a collapsing body of mass $M$ must be given by the following expression:

\begin{equation}
\dot R = - \left( 1 - {2M \over R} \right) ,
\label{edge1}
\end{equation}
whose solution for large enough times is
\begin{equation}
R = 2M + \Delta R_0 e^{-t \over 2M} ,
\end{equation}
with $\Delta R_0$ being a constant depending on the initial conditions of the collapse.

An important detail must be pointed out. In the previous equations we have dealt with the total mass $M$ of the collapsing body as if it was a 
constant. It may be actually considered constant in practice for long periods of time but, in fact, it slowly diminishes due to the emission of 
thermal radiation (unless the surrounding background presents a greater CMB temperature or news amounts of infalling mass are provided). Thus, having into account that $R_S = 2M$, from Eq. \eqref{Hawkingr},

\begin{equation}
\dot M = {-k \over {R_S^2}} =  {-k \over {4M^2}} .
\end{equation}
Therefore,

\begin{equation}
M(t) = \left(M_0^3 -\frac{3k}{4}t \right)^{1 \over 3} ,
\label{masstime}
\end{equation}
from which the evaporation time $t_v$ may be isolated:

\begin{equation}
t_v = {4M_0^3 \over 3k} .
\label{tv} 
\end{equation}

\section{Discussion}
\label{Discussion}
The model of gravitational collapse presented in this paper contains an important number of simplifications which have allowed us to find 
analytical solutions of the coefficients of the metric all over the space at any given time (for small radius values, we have seen that some 
special considerations must be taken into account, but no essential contradiction is risen). The results obtained are self-consistent and do 
not lead to the formation of an event horizon, what would provide a simpler interpretation of the information loss problem: if no event horizon 
is formed, thermal radiation should be directly emitted by the collapsing body. Hence, there is no need for postulating a special mechanism of 
radiation such as the Hawking's one. Let us now analyse more carefully the hypothesis that we have made, their implications and the consequences that would 
have been derived from making slightly different considerations.

Our starting point has been a time-dependent spherically symmetric metric. It is a well-known fact that spherical symmetry is an almost 
universal \emph{approximate} characteristic of any celestial body. Two kind of phenomena certainly prevents it from being \emph{perfect}: the 
first one is rotation (which implies the modification from spherical surfaces to ellipsoidal ones), while the second one consists of the local 
inhomogeneities of any \emph{real} system.

Concerning rotation, it constitutes \emph{per se} a very interesting but mathematically complex problem. To deal properly with a rotating 
process of gravitational collapse, a kind of modified time dependent Kerr metric should be formulated (in the same way that in this paper a 
kind of 'time-dependent Schwarzschild metric' has been proposed). From an intuitive point of view, however, one would expect that rotation 
should lead to a genuinely \emph{slower} collapsing process (due to the 'centrifugal' effect of angular momentum) \cite{Delsate}. Concerning local 
inhomogeneities, a detailed study of the effect of small perturbations on the metric could constitute another \emph{per se} attractive problem, 
but \emph{a priori} it is not unreasonable to assume that the emission of gravitational waves should tend to diminish these effects with time. This is a 
consequence of the 'no hair' theorem for Black Holes (even when we have found no black hole in the mathematical development of this article).

About the temporal dependence of the metric coefficients, it appears to be a strict logical requirement of the problem. The displacement of the 
infalling matter along the collapsing process must necessarily imply a temporal change in the metric coefficients. In this sense, Schwarzschild 
metric -a good solution for the stationary 'punctual mass' problem- is not the best choice for the question of collapse itself. In words of 
J. A. Wheeler, 'matter tells spacetime how to curve, and curved spacetime tells matter how to move'. With our choice of time-dependent metric, 
Kruskal-Szekeres coordinates are not needed because the ordinary polar spherical coordinates cover the entire spacetime manifold and the functions 
$\lambda(r,t)$ and $\nu(r,t)$ are analytic all over the space.

With respect to the choice of stress-momentum tensor, its dust-like nature has been greatly aimed for the sake of simplicity. As it has been already 
emphasized in the pertinent section, it seems paradoxal to consider simultaneously the features of 'dust-like' and 'ultrarelativistic' because the relation 
between pressure and energy density in an ultrarelativistic \emph{gas} turns out to be $p = {1 \over 3} \rho$. Nonetheless, two subtle points should 
be raised here: First of all, the concept of 'ultrarelativistic dust' is not as strange as it appears to be, since a privileged direction of motion 
has been considered (the ultrarelativistic motion is highly 'directed' towards purely radial lines). Secondly, even if a relation of proportionality 
between $p$ and $\rho$ would have been chosen, that would not have changed the fact that all the other stress-momentum tensor components could be 
expressed as a product of certain factors and $T^0_0$. It is straightforward to check that changing the aforementioned factors would not alter drastically 
the subsequent mathematical development. As a matter of fact, the 'linearity' between $T^1_0$ and $T^0_0$ has allowed us to set a temporal dependence 
for $\lambda$. In fact, as $\dot \lambda$ turns out to be proportional to $T^0_0$, the function $\lambda$ would only diverge if $T^0_0$ became infinite too. 
Nevertheless, when $\lambda$ increases $T^0_0$ does not diverge but tends to ${1 \over 8 \pi r^2}$. In a similar way, it may be proved that $\nu$ 
(or $\phi = {- \left( \nu + \lambda \right)/2}$) is also a well-behaved function despite [reasonable] modifications in the stress-momentum tensor.

Thus, whether we consider thermal radiation or not, the study of the temporal evolution of a spherically symmetric gravitational collapse in 
spherical polar coordinates does not lead to incoherences, but constitutes a sensible alternative to the usual Black Hole model. In addition, 
when thermal radiation is considered, very high (but finite) values of $\lambda$ are obtained at any given $r$. 
Definitely, the radiation law proposed in this paper has been deduced in a rather 'heuristic' way by assuming the extensibility of the 
calculations detailed in Ref. \cite{Sto2} to a model of \emph{scarcely interacting} collapsing shells. However, even if the genuine radiation 
law appeared to be completely different, it would still be true that an asymptotic approach to a 'stationary' phase (where the value of $\lambda$ 
does not depend on time) should happen. In fact, this phase should be always reached just by assuming the reasonable hypothesis that the outgoing 
flux of thermal radiation should not diminish with time (the temperature of the collapsing body should be expected to rise, with the progression 
of collapse), while the ingoing flux of collapsing matter should become smaller as the spacetime deformation becomes larger.

In summary, even when several of the assumptions of the model of gravitational collapse proposed in this paper may be considered excessively 
'idealistic', it provides an illustrative description of how a time-dependent metric should be the most logical choice for the study of 
gravitational collapse and that the polar spherical coordinates of an asymptotic observer (a scientific on the Earth, not an astronaut falling 
into a black hole) are sufficient to cover the whole collapsing process. The supposed completion of the collapsing process in a finite proper 
time for a co-mobile observer would never be truly accomplished due to the invariance of causal order for any relativistic system (in a finite 
and lesser proper time, the co-mobile observer would be fully evaporated by the emission of thermal radiation). The astronomic objects already 
\emph{identified} as 'black holes' could equally correspond to 'asymptotically collapsing bodies'. Empirically, few differences would be 
expected. From a theoretical point of view, the latter ones may be obtained in a very natural way from the Einstein field equations and avoid many 
of the paradoxes and illogical aspects of the former ones. Thus, according to Occam's razor, asymptotic collapse should be preferred to black 
holes.

\section{Acknowledgements}

The author acknowledges very specially the help and support of I. L\'{o}pez-Aylagas, without which this paper would have never emerged. 
The author also thanks D. Jou and R. Zarzuela for their encouragement, sensible advices and careful review of the whole work.


\begin{thebibliography}{99}
\bibitem{Sch} Schwarzschild, K. ``\"{U}ber das Gravitationsfeld eines Massenpunktes nach der Einsteinschen Theorie''. Sitzungsber. Preuss. Akad. D. Wiss.: 189-196 (1916).
\bibitem{Bir} Birkhoff, G. D. ``Relativity and Modern Physics''. Cambridge, Massachussets: Harvard University Press (1923).
\bibitem{Opp} Oppenheimer, J. R.; Volkoff, G. M. ``On Massive Neutron Cores''. Physical Review 55 (4): 374-381 (1939).
\bibitem{Whe} Wheeler, J. A.; Ford, K. ``Geons, Black Holes, and Quantum Foam: A Life in Physics''. Publication: Norton. New York (1998).
\bibitem{Haw1} Hawking, S.; Ellis, G. F. R. ``The Large Scale Structure of Space-Time''. Cambridge: Cambridge University Press. (1973).
\bibitem{Bek} Bekenstein, J. D. ``Black holes and entropy''. Phys. Rev. D 7:2333-2346 (1973).
\bibitem{Haw2} Hawking, S. W. ``Black hole explosions?''. Nature 248 (5443): 30 (1974).
\bibitem{Arx} Hawking, S. W. ``Information Preservation and Weather Forecasting for Black Holes.'' arXiv preprint arXiv:1401.5761 (2014).
\bibitem{Pen} Penrose, R. ``Gravitational Collapse: The role of General Relativity'';General Relativity and Gravitation, Vol. 34, No. 7, July (2002).
\bibitem{Sto1} Vachaspati, T.; Stojkovic, D.; Krauss, L. M. ``Observation of incipient black holes and the information loss problem''. Phys. Rev. D 76:024005 (2007).
\bibitem{Sto2} Vachaspati, T.; Stojkovic, D. ``Quantum radiation from quantum gravitational collapse''. Phys. Lett. B 663:107-110 (2008).
\bibitem{Pin} Pi\~{n}ol, M.; L\'{o}pez-Aylagas, I. ``Transition from Established Stationary Vision of Black Holes to Never-Stationary Gravitational Collapse.'' arXiv preprint arXiv:1007.2734 (2010).
\bibitem{Mitra1} Mitra, A, ``The Mass of the Oppenheimer-Snyder black hole.'' arXiv preprint astro-ph/9904163 (1999).
\bibitem{Mitra2} Mitra, A. ``Non-occurrence of trapped surfaces and black holes in spherical gravitational collapse.'' Foundations of Physics Letters 13.6 (2000): 543-579.
\bibitem{Mitra3} Mitra, A. ``Why gravitational contraction must be accompanied by emission of radiation in both Newtonian and Einstein gravity.'' Physical Review D 74.2 (2006): 024010.
\bibitem{Mitra4} Mitra, A. ``Radiation pressure supported stars in Einstein gravity: eternally collapsing objects.'' Monthly Notices of the Royal Astronomical Society 369.1 (2006): 492-496.
\bibitem{Robertson1} Robertson, S.L.; Leiter, D.J. ``Evidence for intrinsic magnetic moments in black hole candidates.'' The Astrophysical Journal 565.1 (2002): 447.
\bibitem{Robertson2} Robertson, S.L.; Leiter, D.J. ``On intrinsic magnetic moments in black hole candidates.'' The Astrophysical Journal Letters 596.2 (2003): L203.
\bibitem{Robertson3} Robertson, S.L.; Leiter, D.J. ``The magnetospheric eternally collapsing object (MECO) model of galactic black hole candidates and active galactic nuclei.'' arXiv preprint astro-ph/0602453 (2006).
\bibitem{Kar} Karczmarek, J.; Maldacena, J.; Strominger, A. ``Black Hole Non-Formation in the Matrix Model''. Journal of High Energy Physics. 2006(01), 039.
\bibitem{Lan} Landau, L. D.; Lifshitz, E. M. ``The classical  theory of Fields''. Course of Theoretical Physics  Volume 2. University of Minnesota (1987).
\bibitem{Yan} Lu, Y. ``Black Hole Radiation and Energy-Momentum Tensor''. Diploma Thesis for Theoretical Physics in Utrecht University. Netherlands (2010).
\bibitem{Haz} Hazlehurst, J.; W. L. W. Sargent. ``Hydrodynamics in a Radiation Field-A Covariant Treatment.'' The Astrophysical Journal 130 (1959): 276.
\bibitem{Bor} Borkar, M. S.; Dhongle, P. R. ``Pre-Hawking Radiating Gravitational Collapse in Stationary Space-Time.'' International Journal of Theoretical and Applied Sciences, 5(2): 27-31 (2013).
\bibitem{Delsate} Delsate, T.; Rocha, J.V.; Santarelli, R. ``Collapsing thin shells with rotation.'' arXiv preprint arXiv:1405.1433 (2014).
\end{thebibliography}
\end{document}